\input phyzzx.tex
\tolerance=1000
\voffset=-0.0cm
\hoffset=0.7cm
\sequentialequations
\def\rl{\rightline}

\def\t1{{\tilde 1}}

\def\t{\theta}

\REF{\BEK}{J. Bekenstein, Lett. Nuov. Cimento {\bf 4} (1972) 737; Phys Rev. {\bf D7} (1973) 2333; Phys. Rev. {\bf D9} (1974) 3292.}
\REF{\HAW}{S. Hawking, Nature {\bf 248} (1974) 30; Comm. Math. Phys. {\bf 43} (1975) 199.}
\REF{\HOL}{G. 't Hooft, gr-qc/9310026; L. Susskind, J. Math. Phys. {\bf 36} (1995) 6377, hep-th/9409089.}
\REF{\RAP}{R. Bousso, JHEP {\bf 9907} (1999) 004, hep-th/9905177; JHEP {\bf 9906} (1999) 028, hep-th/9906022; JHEP {\bf 0104} (2001) 035, hep-th/0012052.}
\REF{\VAF}{A. Strominger and C. Vafa, Phys. Lett. {\bf B379} (1996) 99, hep-th/9601029.}
\REF{\CAL}{C. Callan and J. Maldacena, Nucl. Phys. {\bf B472} (1996) 591, hep-th/9602043; G. Horowitz and A. Strominger, Phys. Rev. Lett. {\bf 77} (1996)
2368, hep-th/9602051; J. Maldacena and A. Strominger, Nucl. Phys. {\bf B472} (1996) 591, hep-th/9602043; G. Horowitz, D. Lowe and J. Maldacena, Phys. Rev.
Lett. {\bf 77} (1996) 430; hep-th/9603195.}
\REF{\HOR}{G. Horowitz, hep-th/9704072 and references therein.}
\REF{\LEN}{L. Susskind, hep-th/9309145.}
\REF{\SBH}{E. Halyo, A. Rajaraman and L. Susskind, Phys. Lett. {\bf B392} (1997) 319, hep-th/9605112.}
\REF{\HRS}{E. Halyo, B. Kol, A. Rajaraman and L. Susskind, Phys. Lett. {\bf B401} (1997) 15, hep-th/9609075.}
\REF{\HP}{G. Horowitz and J. Polchinski, Phys. Rev. {\bf D55} (19997) 6189, hep-th/9612146.}
\REF{\HPO}{G. Horowitz and J. Polchinski, Phys. Rev. {\bf D57} (1998) 2557, hep-th/9707170.}
\REF{\DAM}{T. Damour and G. Veneziano, Nucl. Phys. {\bf B568} (2000) 93, hep-th/9907030.}
\REF{\JUA}{J. Maldacena, Nucl. Phys. {\bf B477} (1996) 168, hep-th/9605016.}
\REF{\EDI}{E. Halyo, Int. Journ. Mod. Phys. {\bf A14} (1999) 3831, hep-th/9610068; Mod. Phys. Lett. {\bf A13} (1998), hep-th/9611175.}
\REF{\DAN}{U. Danielsson, A. Guijosa and M. Kruczenski, hep-th/0106201.}
\REF{\DES}{E. Halyo, hep-th/0107169.}
\REF{\HST}{G. Horowitz and A. Strominger, Nucl. Phys. {\bf B360} (1991) 197.}
\REF{\BTZ}{M. Banados, C. Teitelboim and J. Zanelli, Phys. Rev. Lett. {\bf 69} (1992) 1849; M. Banados, M. Henneaux, C. Teitelboim and J. Zanelli, Phys.
Rev. {\bf D48} (1993) 1506.}
\REF{\KLE}{I. R. Klebanov and A.A. Tseytlin, Nucl. Phys. {\bf B475} (1996) 164, hep-th/9604089.}
\REF{\DUF}{M. J. Duff, H. Lu and C.N. Pope, Phys. Lett. {\bf B382} (1996) 73, hep-th/9604052.}
\REF{\CON}{J. D. Brown and M. Henneaux, Commun. Math. Phys. {\bf 104} (1986) 207.}
\REF{\STR}{A. Strominger, JHEP {\bf 02} (1998) 009.}
\REF{\CAR}{S. Carlip, Phys. Rev. Lett. {\bf 82} (1999) 2828, hep-th/9812013.}
\REF{\SOL}{S. Solodukhin, Phys. Lett. {\bf B454} (1999) 213.}

\singlespace
\rl{SU-ITP-01-}
\rl{hep-ph/0108167}
\rl{\today}
\pagenumber=0
\normalspace
\medskip
\bigskip
\titlestyle{\bf{Universal Counting of Black Hole Entropy by Strings on the Stretched Horizon}}
\smallskip
\author{ Edi Halyo{\footnote*{e--mail address: vhalyo@.stanford.edu}}}
\smallskip
\centerline {Department of Physics}
\centerline{Stanford University}
\centerline {Stanford, CA 94305}
\centerline{and}
 \centerline{California Institute for Physics and Astrophysics}
\centerline{366 Cambridge St.}
\centerline{Palo Alto, CA 94306}
\smallskip
\vskip 2 cm
\titlestyle{\bf ABSTRACT}

We show that the entropy of any black object in any dimension can be understood as the entropy of a highly excited string on the stretched horizon.
The string has a gravitationally renormalized tension due to the large redshift near the horizon. The Hawking temperature is given by the Hagedorn
temperature of the string. As examples, we consider black holes with one (black p--branes) or more charges, Reissner--Nordstrom black holes and the BTZ black hole
in addition to Schwarzschild black holes.
We show that the vanishing and nonvanishing extremal entropies can be obtained as smooth limits of the near--extreme cases.

\singlespace
\vskip 0.5cm
\endpage
\normalspace

\centerline{\bf 1. Introduction}
\medskip

The microscopic origin of the Bekenstein--Hawking formula[\BEK,\HAW] which relates gravitational entropy to horizon area
$$S_{BH}={A_{hor} \over {4G}} \eqno(1)$$
has always been mysterious.
This entropy formula is universal; it applies to all black objects with a horizon. Therefore it is a hint about the fundamental nature of
quantum gravity and/or horizons. The most important lesson to be learned from (1) may be holography[\HOL, \RAP], i.e. that gravity in D dimensions can be described by a
nongravitational theory in one less dimension with one degree of freedom per Planck area.
The universality of the Bekenstein--Hawking entropy formula leads us to believe that it must have a universal microscopic explanation
in quantum gravity. Since string theory is the only consistent theory of quantum gravity we should look for such an explanation in string theory.

During the last few years, great progress was made in microscopic counting of black hole entropy in string theory[\VAF,\CAL,\HOR].
Microscopic entropy of many black holes made of
D--branes was explained in terms of the degrees of freedom that live on the branes.
However, this was accomplished mainly for
extreme and near-extreme black holes (and nondilatonic black p--branes) and used supersymmetry in a crucial way. Unfortunately, the same methods cannot be used
for generic black holes or black p--branes far from extremality since the power of supersymmetry is lost in these cases.

For general black holes (and horizons) there has to be a different entropy counting than the one used for the supersymmetric D--brane systems. Since there is no
supersymmetry, any such counting can only be approximate, i.e. we can only expect to get the correct scaling of the entropy without the exact coefficients.
In any case, there may be large corrections to such coefficients since black holes are also strongly interacting systems.
In ref. [\LEN,\SBH] the entropy of a $D$ dimensional Schwarzschild black hole was obtained by viewing it as a highly excited string with a gravitationally
renormalized tension. The prescription used in ref. [\LEN,\SBH] is as follows. The near horizon geometry of a (e.g. four dimensional)  Schwarzschild black
hole is Rindler space with a dimensionless time $\tau_R$
and energy $E_R \sim GM^2$. We see that $E_R \sim S$ which is a macroscopic relation obtained from the geometry. In order to understand the microscopic reason
behind this relation
consider a string with tension $T= 1/ 2\pi \ell_s^2$ living on the stretched horizon at $r=r_0+\ell_s$. This string has energy
$E_{sh} \sim E_R/\ell_s$
and therefore is in a very highly excited state with oscillator number $n \sim E_R^2$. Then, the entropy of the string is $S \sim \sqrt n \sim E_R \sim GM^2$ which gives the
entropy of the black hole. On the other hand, an asymptotic observer sees a gravitationally renormalized string energy (mass) and tension due to the large redshift
near the horizon. The asymptotic energy of the string is $E_{as} \sim E_R/GM \sim M$ i.e. the ADM mass of the black hole whereas its tension is renormalized from
$T= 1/ 2\pi \ell_s^2$ to
$T \sim (GM)^{-2}$. The entropy of the string which is a number remains the same. The Hawking temperature of the SBH is given by the Hagedorn temperature of the string
with the renormalized tension, i.e. $T_H \sim T_{Hag} \sim 1/GM$.
This result can be generalized to Schwarzschild black holes in all dimensions.
Thus, the entropy of a Schwarzschild black hole due to the horizon can be understood as the entropy
of a highly excited string on the stretched horizon with a gravitationally rescaled tension.
It has been noticed that the above prescription, in addition to Schwarzschild black holes[\HRS,\HP,\HPO,\DAM]
works in other cases where the entropy of the system can be described by strings, e.g. near extreme
$D5$ branes [\JUA], some near extremal $D=4,5$ black holes[\EDI] and brane--antibrane systems[\DAN]. Recently, it was shown that the same prescription can also
be used to explain the entropy of (the static patch of) de Sitter space in any dimension[\DES].

On the other hand, the identification of the Rindler energy with entropy seems to be very general. For any metric with a nonsingular horizon, the near horizon
geometry is described by Rindler space. In (Euclidean) Rindler space we can write the first law of thermodynamics as $dE_R=T_R dS$. $T_R$ is the dimensionless
Rindler temperature which is given by the periodicity of the Euclidean Rindler time, $T_R=1/2 \pi$. Thus we find that $S=2 \pi E_R$ and the entropy is always
given by the Rindler energy. This result can be explained by assuming the existence of a string on the stretched horizon as above even for cases in which there
are no fundamental closed strings.

In this paper, we show that the prescription of ref. [\LEN,\SBH] is much more general and can be applied to any black object with gravitational entropy.
This includes black p--branes, black holes with two or more charges, $D=4,5$ Reissner--Nordstrom black holes and the BTZ black hole (in addition to
Schwarzschild black holes and de Sitter spaces in any dimension that already appear in the literature).
The gravitational entropy of any black object can be understood as that of a highly excited string on the stretched
horizon with a renormalized
tension. In every case, the Hawking temperature is given by the Hagedorn temperature of the string with the renormalized tension. The string carries the nonextremal
part of the mass of the black object and its total entropy ($D=4,5$ Reissner--Nordstrom black holes are exceptions which we deal with in section 4).
In each case the metric has a Schwarzschild--like factor which gives rise to the horizon. The near
horizon geometry is Rindler space which allows the prescription of ref. [\LEN,\SBH] to be used. We show explicitly that the above method gives the correct
entropy for all black p--branes in any dimension[\HST], dyonic strings, M--branes, $D=4$ Reissner--Nordstrom black holes and BTZ black holes in $D=3$[\BTZ].
The (vanishing) entropy of extreme p--branes are obtained as smooth extremal limits of the nonextremal entropy. $D=4,5$ Reissner--Nordstrom black holes are
special cases because they have nonzero extreme entropy. We show that their (extreme and nonextreme) entropy can also be given by that of a string on the
stretched horizon if the string carries two of the black hole charges.
For all black objects we obtain the correct scaling of the entropy both in the near extreme and the highly
nonextreme limits. For technical reasons it is harder to show this in general. Since the black hole entropy and the string tension are continous functions
of the nonextremality parameter, we assume that our results hold for any amount of nonextremality.
The fact that our prescription works for all black objects irrespective of their composition leads us to believe that the strings on the
stretched horizon with renormalized tension describe the fundamental degrees of freedom of the horizon.

The paper is organized as follows. In the next section we show that our prescription for entropy counting applies to the most general
solutions of supergravity which describe black objects. In section 3 we explicitly consider a few examples such as 3 and 5--branes, the dyonic string,
M2 branes and the Schwarzschild limit.
In section 4, we describe the $D=4$ Reissner--Nordstrom black holes which require a slightly different approach. In section 5, we discuss the $D=3$ BTZ black holes.
Section 6 contains a discussion of our results and our conclusions.

\medskip
\centerline{\bf 2. Gravitational Entropy and Strings on the Stretched Horizon}
\medskip

In this section we find the entropy and temperature for the most general black object using our prescription and show that they agree with results obtained
in supergravity. We show that the entropy of all black objects with nonsingular horizons can be understood as the entropy of a string on the
stetched horizon with a renormalized tension. The Hawking temperature is given by the Hagedorn temperature of the string with renormalized tension.
We find that these results hold for a wide range of nonextremality, from the extreme to the highly nonextreme cases including the Schwarzschild limit.

We first show that the prescription of ref. [\LEN,\SBH] applies to all black objects (p--branes and black holes) in any dimension.
Black objects are extrema
of the D--dimensional dilatonic supergravity action
$$S=-{1 \over \kappa^2} \int d^Dx \sqrt{g}[R-{1 \over 2}(\partial \phi)^2-{1 \over{2(d+1)!}} e^{a \phi} F^2_{d+1}] \eqno(2)$$
where $D=p+d+3$, $\kappa^2/8 \pi$ is the D--dimensional Newton constant, $a$ is a parameter and $\phi$ is the dilaton.
The most general solution of the above action  for a black object is given by[\KLE,\DUF]
$$ds^2=H^{\alpha}(H^{-N}[-f(r)dt^2+dy_1^2+ \ldots +dy_p^2]+f(r)^{-1}dr^2+r^2d \Omega^2_{d+1}) \eqno(3)$$
where
$$H(r)=1+{r_0^d \over r^d}, \qquad f(r)=1-{\mu^d \over r^d}, \qquad  r_0^d=\mu^d sinh^2 \gamma  \eqno(4)$$
In addition, the solutions involve the field strength $F_{d+1}$ and the dilaton $\phi$.
Note that the above metric is the Einstein metric and not the string metric $ds^2_{st}=H^{(3-p)/12}(r) ds^2_E$ that the string sees. However, we will see that overall
factors that multiply the metric (such as $H^{\alpha}(r)$) drop out of our calculation so using the Einstein metric gives the same answer as using the string metric.
This is as it should be since the entropy is the same in the string and Einstein
frames. The parameters $\alpha$ and $N$ (which gives the number of charges carried by the black hole) are given by
$$\alpha={N(p+1) \over {D-2}}, \qquad N=4[a^2+{2d(p+1) \over {D-2}}]^{-1} \eqno(5)$$
specific values of $D$, $d$ and $N$ describe different black objects.
For example, black p--branes correspond to the solution with $N=1$ and $D=10$ whereas a dyonic
self--dual string is given by $p=1$, $N=2$ and $D=6$. On the other hand, Reissner--Nordstrom black holes in $D=4,5$ correspond to solutions with $d=1$, $N=4$ and $d=2$, $N=3$
(and with $D=10$) respectively. $D$ dimensional Schwarzschild black holes are obtained by substituting $p=0$ and $N=0$ in eq. (3).
The extremal limit is obtained by $\mu \to 0$ and $\gamma \to \infty$ with $\mu^d sinh^2 \gamma$ fixed.
We assume that the internal dimensions of the black object, $y_i$ are compactified on a torus of radius $L$. The charge per unit volume is
$$q_p={\omega_{d+1} \over {2 \sqrt{2} \kappa}} \sqrt N d \mu^d sinh (2 \gamma)={\omega_{d+1} \over { \sqrt{2} \kappa}} \sqrt N d r_0^d \eqno(6)$$
where $\omega_{d+1}$ is the volume of the unit $d+1$ sphere. The ADM mass is given by
$$M_p={\omega_{d+1} \over {2  \kappa^2}} L^p \mu^d (d+1+Nd sinh^2 \gamma) \eqno(7)$$
The $D-2$ dimensional area of the horizon located at $r=\mu$ is
$$A_p=\omega_{d+1} L^p \mu^{d+1} H^{N/2}(\mu) \eqno(8)$$
The Bekenstein--Hawking entropy of the black object is then $S_p=2 \pi A_p/\kappa^2$.
In the near extreme limit this gives
$$S \sim {\omega_{d+1} L^p \over \kappa^2} r_0^{dN/2} \mu^{d+1-dN/2}  \eqno(9)$$
whereas in the opposite limit of a highly nonextremal black object we have
$$S \sim  {\omega_{d+1} L^p \over \kappa^2} \mu^{d+1} \eqno(10)$$
The Hawking temperature for the black object is
$$T_H= \omega_{d+1}^{1/d} \kappa^{N/2-2/d} d^{-(1+1/d)} \lambda^{1-\lambda} L^{p(1-\lambda)} q_p^{-N/2} E^{(1-\lambda)} \eqno(11)$$
where $E$ is the deviation of the ADM mass from its extremal value and $\lambda=(1+1/d)-N/2$.

Let us now apply our prescription to the generic solution above. Note that like the Schwarzschild and de Sitter cases the metric in eq. (3) has the factor
$1-\mu^d/r^d$ which means that the near horizon geometry is Rindler space and the prescription of ref. [\LEN,\SBH] can be used. Near the horizon, we have $r=\mu+y$ with
$y<<\mu$ and the metric becomes
$$ds^2=H^{\alpha}(\mu)(H^{-N}(\mu)[-(dy/\mu)dt^2+dy_1^2+ \ldots +dy_p^2]+(dy/\mu)^{-1}dy^2+\mu^2d \Omega^2_{d+1}) \eqno(12)$$
The proper distance to the horizon is
$$R= \sqrt{\mu \over d} H^{\alpha/2}(\mu) \sqrt y \eqno(13)$$
Then the near horizon metric becomes
$$ds^2=(H^{-N}(\mu)[-(R^2 d^2/\mu^2)dt^2+\ldots]+dR^2+\ldots) \eqno(14)$$
which is the metric for Rindler space in the $t-r$ plane.
The dimensionless Rindler time is
$$\tau_R \sim H^{-N/2}(\mu) {t \over \mu} \eqno(15)$$
The dimensionless Rindler energy is conjugate to $\tau_R$
$$[E_R,\tau_R]=1=[E_R,t] {H^{-N/2}(\mu) \over \mu} \eqno(16)$$
The deviation of the energy from extremality for $\mu<<r_0$ or $\mu>>r_0$ is obtained from eq. (7) to be
$$E \sim {\omega_{d+1} L^p \over \kappa^2} \mu^d \eqno(17)$$
Using
$[E_R,t]=\partial E_R/\partial E$ we find (for $\mu<<r_0$ or $\mu>>r_0$)
$$dE_R \sim {\omega_{d+1} L^p \over \kappa^2} \int H^{N/2}(\mu) \mu^d d\mu \eqno(18)$$
In the near extreme limit $\mu << r_0$ we get the Rindler energy
$$E_R \sim {\omega_{d+1} L^p \over \kappa^2} r_0^{dN/2} \mu^{d+1-dN/2}  \eqno(19)$$
We see that (in the near extremal limit $\mu<<r_0$) the Rindler energy $E_R$ is the entropy of the black object given by eq. (9) for any $p, N, D$.
In the highly nonextreme limit $\mu>> r_0$ we find
$$E_R \sim  {\omega_{d+1} L^p \over \kappa^2} \mu^{d+1} \eqno(20)$$
which is again the entropy in eq. (10).

It is much harder to obtain $E_R$ for the intermediate case with $\mu \sim r_0$ for two reasons. First, the expression for $E$ is in general complicated. Second,
the factor of $H^{N/2}(\mu)$ in eq. (18) for $E_R$ leads to a very complicated integral unless we take one of the above limits. We would like to argue that since we
$S \sim E_R$ both in the near--extreme and highly nonextreme limits and the string and black hole entropies are continous functions of the mass
the identification should also hold for $\mu \sim r_0$. It would be nice to show this explicitly for any $\mu$ and $r_0$.

The connection between the entropy and $E_R$ can be understood if we consider a string on the stretched horizon at $r=\mu+\ell_s$.
The energy of the string with tension $T=1/2\pi \ell_s^2$ is
$$E_{sh}\sim {E_R \over \ell_s} \sim {\omega_{d+1} L^p \over {\ell_s \kappa^2}} r_0^p \mu^r \eqno(21)$$
where $r=dN/2$, $p=d+1-dN/2$ for $\mu<<r_0$ and $r=0$, $p=d+1$ for $\mu>>r_0$.
We find that for this string $E_R \sim \sqrt n$ and therefore the string entropy $S= 2 \pi \sqrt {cn/6}$ gives the entropy of the black object $S_p$.
An asymptotic observer at infinity sees a gravitationally redshifted string energy and tension (but the same entropy which does not get redshifted since it is a number).
This is a large effect due to the large redshift near the horizon.
The string mass seen from infinity is in both limits (using eq. (15))
$$M \sim {E_R \over {\mu H^{N/2}(\mu)}} \sim {\omega_{d+1} \over \kappa^2} L^p \mu^d \eqno(22)$$
which is the deviation of the ADM mass of the black object from its value at extremality given by eq. (7).
(We note in this connection that the extreme part of the mass does not get gravitationally renormalized. We do not have an explanation for this since the
overall supersymmetry is broken.)
Using the equation for the string mass we find the gravitationally renormalized tension of the string seen by asymptotic observers
to be
$$T \sim {1 \over {\mu^2 H^N(\mu)}} \eqno(23)$$
We see that the renormalized tension of the string is proportional to the deviation from extremality and inversely proportional to the charge (for $\mu<< r_0$).
Therefore, the renormalized tension decreases as the configuration gets closer to the extreme one. For near--extreme black objects with $\mu<<r_0$ we find
$T \sim \mu^{dN-2}/r_0^{dN}$; therefore the tension seen by an asymptotic observer is extremely small if $r_0>>\mu$.
On the other hand, for $\mu>>r_0$ the tension decreases as
$\mu$, i.e. as the nonextremality increases. We find $T \sim 1/\mu^2$ which is again very small for $\mu>>r_0$.
The Hawking temperature of the black object is given by the Hagedorn temperature of the string with the above tension, i.e.
$$T_H \sim T_{Hag} \sim {1 \over {2 \pi \mu H^{N/2}(\mu)}} \eqno(24)$$
The same result can also be obtained by using eq. (15) and noting that $T_R=1/ 2 \pi$.

We interpret these results as follows. The entropy of any black object in any dimension can be understood as the entropy of a highly excited
string on the stretched horizon.
The string tension (seen by an asymptotic observer) is gravitationally renormalized to a very small value due to the large gravitational redshift near the horizon.
The Hawking temperature of the black object is given by
the Hagedorn temperature of the string with the renormalized tension.
The generality of our results and their lack of dependence on the composition of the black objetcs and/or the microscopic degrees of freedom
seem to indicate that the string
describes the fundamental degrees of freedom on the horizon which count the horizon entropy. The highly excited string on the stretched horizon
is very long, with length $\sim E_R \ell_s$.
Its large entropy proportional to its total length can be explained if we assume that it is made of string bits of length $\ell_s$ which carry one bit of information.
This suggests that the fundamental degrees of freedom on the horizon are the string bits of length $\ell_s$.

We have described the entropy of black objects with nonsingular horizons in terms of a string on the stretched horizon. What is the description for extreme
black holes and p--branes? The near horizon geometry for these is not Rindler space so it seems that our prescription is not suitable for these cases.
We will see in the next section that the vanishing entropy of the extreme black objects can be obtained as smooth limits of the nonextreme entropies in terms of a string as above.
In these cases, as we take the extreme limit both the mass and entropy of the string vanish. As we mentioned, a long closed string cannot account for the entropy of
Reissner--Nordstrom black holes which have nonzero extreme entropy. We show in section 4 that for these black holes the string on the stretched horizon must carry
two of the black hole charges.

The nature of the string on the
stretched horizon is not clear to us. There are indications that it may be a noncritical string with $c=6$[\HRS]. The change in the central charge may be interpreted as
gravitational renormalization. More importantly, we used the perturbative formulas for the string energy and
entropy even though the string on the stretched horizon should be strongly interacting due to its large number of degrees of freedom. In our picture,
somehow this long string is behaving as if it is free.
We are not able to explain this fact other than to note that it works in all cases. Note that for a string $S \sim E \ell_s$ so that
our results do not change if there are strong coupling effects which change the entropy and energy by a proportional amount. However we cannot show this for strongly
interacting strings.

\medskip
\centerline{\bf 3. Examples}
\medskip

In this section we consider a few examples of black objects explicitly in order to gain a better understanding of the strings on the stretched horizon. We first
discuss the best understood black p--branes which are 3 and 5--branes and show the connection between the world--volume picture and ours. We then consider
dyonic strings as examples with more than one charge. We show that the highly nonextreme limit of p--branes agrees with the Schwarzschild limit. Finally we
show that our prescription works for M2 branes even though in this case there cannot be a microscopic explanation by a string on the stretched horizon.

{\bf 3--branes:} For near--extremal 3--branes the energy beyond extremality is $E_3 \sim \omega_5 L^3 \mu^4/\kappa^2$ and the renormalized tension of the string is given by
$T \sim \mu^2/r_0^4$. The entropy and Hawking temperature are
$$S_3 \sim E_R \sim {\omega_5 L^3 \over  \kappa^2} \mu^3 r_0^2 \qquad \qquad T_H \sim \sqrt T \sim {\mu \over r_0^2} \eqno(25)$$
respectively. We see that the entropy can be written as
$S_3 \sim E^{3/4} L^{3/4}$ which is
exactly the relation we would get from considering the entropy of a four dimensional noninteracting gas[\KLE]. The temperature of the gas is found to be $T_H$. This is the
microscopic explanation of the entropy of near--extreme D3 branes using the world--volume degrees of freedom.
The above description in terms of a string is valid for $\mu<<r_0$ which means for energies $E_3 << N/\ell_s$ where $N$ is the number of 3--branes. For $N>>1$
we see that the deviation from extremality can be much larger than the string mass. This should be contrasted with the world--volume methods which are only
valid up to energies $\sim 1/\ell_s$.

{\bf 5--branes:} For near--extremal 5--branes the energy beyond extremality is $E_5 \sim \omega_3 L^5 \mu^2/\kappa^2$ with the renormalized tension of the
string $T \sim 1/r_0^2$.
The entropy and Hawking temperature are given by
$$S_5 \sim E_R \sim {\omega_3 L^5 \over  \kappa^2} \mu^2 r_0 \qquad \qquad  T_H \sim \sqrt T \sim {1 \over r_0} \eqno(26)$$
respectively.
In this case, the entropy can be written as $S_5 \sim E_5 r_0$
which is the relation for a string with fractional tension $1/r_0^2$ living on the 5--brane. This is exactly the microscopic description of near--extreme
5--brane entropy given in ref. [\JUA].

In both of the above cases the extreme p--brane entropy vanishes. We see that as the deviation from extremality $\mu \to 0$ the nonextreme energy $E$ goes to zero
faster than $T_{Hag}$. Therefore $n \to 0$ as $\mu \to 0$ and the entropy vanishes.
However, there are cases like $D=4,5$ Reissner--Nordstrom black holes with nonzero entropy at
extremality. These black holes do not have Rindler space as the near horizon geometry.
In the next section we show that our prescription also works for these cases if we start with the near extremal black holes and take the extremal limit which is smooth.

{\bf Dyonic strings:} As an example of a black hole with two charges we consider the $D=6$ dyonic (self--dual) string on $S^1$ which is a $D=5$ black hole with two
charges. This corresponds to the solution in eq. (3) with $D=6$, $d=2$ and $N=2$. Then, we find the energy beyond extremality to be $E_2 \sim \omega_3 L \mu^2/\kappa^2$.
The renormalized tension of the string is given by $T \sim \mu^2/r_0^4$ and therefore the Hawking temperature is $T_H \sim \mu/r_0^2$. As a result, the entropy
is $S_2 \sim \omega_3 L r_0^2 \mu/\kappa^2$. We see that the entropy can be written as $S_2 \sim (E L)^{1/2}$ which is the relation for a gas in two dimensions.

{\bf The highly nonextreme limit:} We can also go way beyond the near--extreme cases by taking $\mu>>r_0$. This corresponds to Schwarzchild
black holes which were considered in ref. [\LEN,\SBH]. Then in $D$
dimensions we have (using $\mu^d=GM$ and $d=D-3$)
$E \sim \omega_{d+1} \mu^d/\kappa^2 \sim M$ with a string tension $T \sim 1/\mu^2 \sim 1/(GM)^{2/(D-3)}$. The Hagedorn temperature of the string gives the Hawking
temperature $T_H \sim 1/\mu \sim  1/(GM)^{1/(D-3)}$. The entropy is
$$S \sim {\omega_{d+1} \mu^{d+1} \over \kappa^2} \sim G^{1/(D-3)} M^{(D-2)/(D-3)} \eqno(27)$$

All black p--branes reduce to Schwarzschild black holes in this limit (when their world--volume is compactified). Consider for example the p--brane on $T^p$ with radii $L$.
For $\mu>>r_0$ the energy is $E_p \sim \omega_{d+1} \mu^d/\kappa_{D-p}^2$ where $\kappa_{D-p}^2=\kappa_D^2/L^p$.
The entropy is $S_p \sim \omega_{d+1} \mu^{d+1}/\kappa_{D-p}^2$. We can now write $S_p \sim E_p^{d+1/d}$ which is eq. (10) for $d=D-3$.

{\bf M--branes:} The solution for M2 and M5 branes are given by eq. (3) with $N=1$, $D=11$ and $d=6$ and $d=3$ respectively. The identification
$S \sim E_R$ is also valid for M--branes. However, in M theory there are no
strings so our microscopic description in terms of a string does not make sense. The only scale in the theory is the fundamental scale $\ell_{11}$ so
if there were a stretched horizon it
would be a distance $\ell_{11}$ away from the event horizon. In addition, one would need an M theory object with entropy $S \sim E \ell_{11}$.
Here we consider only the
M2 brane since the M5 brane is similar. The M2 brane energy beyond extremality is given by $E_2 \sim \omega_7 L^2 \mu^6/\kappa^2$. The renormalized string tension
is $T \sim \mu^4/r_0^6$ which gives the entropy
$$S_2 \sim E_R \sim {\omega_7 L^2 \over \kappa^2} \mu^4 r_0^3 \eqno(28)$$
which agrees with the supergravity result. Note that we can write the entropy as $S_2 \sim E_2^{2/3} L^{2/3}$ which is the relation that describes a three
dimensional free gas. The temperature of the gas is found to be $T_H$.

\medskip
\centerline{\bf 4. Reissner--Nordstrom Black Holes}
\medskip

As we saw above the entropy of any black object can be understood as the entropy of a highly excited string on the stretched horizon. The string carries the
nonextremal part of the energy and all the entropy of the black object. Therefore, it is clear that it cannot describe the entropy of Reissner--Nordstrom
black holes in $D=4,5$ which have nonzero entropy at extremality. At extremality the long closed string has to be massless and therefore cannot carry
any entropy. A possible solution is for the string to carry  some charges and part of the extreme mass of the black hole. Then, at extremality
the string is in a BPS state and can have nonzero entropy. This is in fact the approach taken in ref. [\EDI]. Note that since the string carries two charges,
the Reissner--Nordstrom
black hole must have at least two NS--NS charges. Black holes with Ramond--Ramond charges can be obtained from these by dualities.

We now consider the $D=4$ Reisssner--Nordstrom black hole with at least two NS--NS charges, e.g. winding and momentum along one of the compact dimensions carried by
the string on the stretched horizon. (The $D=5$ Reissner--Nordstrom black hole with at least two NS--NS charges is very similar.)
The string
is in the background of the metric in eq. (3) but now with $N=2$ (instead of $N=4$) and $D=4$, $d=1$. The black hole mass is
$$M_{RN} \sim {\omega_2 \over \kappa^2} (r_0+\mu) \eqno(29)$$
When the four charges of the black hole are comparable and the deviation from extremality $\mu$ is carried only by the string, the string mass is
$M_{str} \sim M_{RN}$. Using eq. (16) the Rindler energy is given by
$$E_R \sim {\omega_2 \over \kappa^2} (r_0+\mu)^2 \eqno(30)$$
We see that this agrees with the entropy $S_{RN}$ obtained from eq. (8) with $D=4$, $N=4$ and $d=1$. We now have a string on the stretched horizon with energy
$$E_{sh} \sim {\omega_2 \over {\kappa^2 \ell_s}} (r_0+\mu)^2 \eqno(31)$$
and tension $T =1/2 \pi \ell_s^2$. It is in a slightly non--BPS state with left and right oscillator numbers
$N_L \sim \omega^2_2 r_0^4/\kappa^4>>1$ and $N_R \sim \omega^2_2 (\mu^2+\mu r_0)^2/\kappa^4<<1$.
Using the metric (with $N=2$) we find the Rindler time
$$\tau_R \sim {t \over (r_0+\mu)} \eqno(32)$$
This means that an asymptotic observer sees a string mass
$$E_{str} \sim {\omega_2 \over \kappa^2 } (r_0+\mu) \eqno(33)$$
which equals $M_{RN}$. The tension of the string is gravitationally renormalized to
$$T \sim {1\over (r_0+\mu)^2} \eqno(34)$$
The Hawking temperature of the black hole is $T_H \sim \sqrt T \sim 1/2 \pi (r_0+\mu)$ which agrees with eq. (9).

We see that the above description has a smooth exterme limit for $\mu \to 0$ in which $S_{RN} \sim \omega_2 r_0^2 / \kappa^2$. Now the string on the stretched horizon
is in a BPS state with $N_L \sim \omega^2_2 r_0^4/\kappa^4>>1$ and $N_R=0$. The string tension seen from infinity becomes $T \sim 1/2 \pi r_0^2$.
Note that $S_{RN}$ given by eq. (30) goes over from the extreme case to the Schwarzschild case smoothly as $\mu$ becomes larger compared to $r_0$. During this
transition the renormalized tension of the string goes from $T \sim 1/(\mu+r)^2$ to $T \sim 1/r_0^2$ smoothly.

Above we described the $D=4$ Reissner--Nordstrom black hole by a string (with two charges) in the background of a black hole with two charges.
This should not be confused
with the description of a black hole with two charges. In that case the entropy is carried by a string on the stretched horizon which does not carry any charge;
therefore the extreme entropy of a black hole with two charges vanishes

\medskip
\centerline{\bf 5. BTZ Black Holes}
\medskip

The black hole solutions given in eq. (3) all have an asymptotically flat metric.
As an example which is qualitatively different from these we consider the $2+1$ dimensional BTZ black hole which does not have an asymptotically flat metric.
The BTZ black hole is described by the metric[\BTZ]
$$ds^2=-(-4MG+{r^2 \over L^2})dt^2+(-4MG+{r^2 \over L^2})^{-1}dr^2+r^2d\phi^2 \eqno(35)$$
which is a solution of $2+1$ gravity with a negative cosmological constant $\Lambda=-1/L^2$ with the action
$$I={1 \over {16 \pi G}} \int d^3x \sqrt{-g}(R+2\Lambda) \eqno(36)$$
There is a horizon at $r=r_0=(4GML^2)^{1/2}$
and the entropy is given by
$$S_{BTZ}={2 \pi r_0 \over G} = 4 \pi L(M/G)^{1/2} \eqno(37)$$
The Hawking temperature is
$$T_{BTZ}={(MG)^{1/2} \over {2 \pi L}} \eqno(38)$$
We now use our prescription and go to the near horizon limit by taking $r=r_0+y$ with $y<<r_0$. Then the metric becomes
$$ds^2=-({2yr_0 \over L^2})dt^2+({2yr_0 \over L^2})^{-1}dy^2+r_0^2d\phi^2 \eqno(39)$$
The proper distance to the horizon is given by
$$R=\sqrt{1 \over 2 r_0} L \sqrt y \eqno(40)$$
and the metric takes the Rindler form
$$ds^2=-({4 r_0^2 \over L^4})R^2dt^2+dR^2+r_0^2 d\phi^2 \eqno(41)$$
We see that the Rindler time is
$$\tau_R \sim {r_0 \over L^2} t \eqno(42)$$
The dimensionless Rindler energy satisfies $[E_R,\tau_R]=1=[E_R,t](r_0/L^2)$. This means
$$dE_R={L^2 \over r_0}dM={L^2 \over {(4GL^2)^{1/2}}}{dM \over M^{1/2}} \eqno(43)$$
We find the Rindler energy $E_R \sim L(M/G)^{1/2}$.
Again if we consider a string (with tension $T=1/2 \pi \ell_s^2$) on the stretched horizon at $r=r_0+\ell_s$ then its energy would be
$E_{sh} \sim E_R/\ell_s$ and therefore the string oscillator number is
$\sqrt n \sim E_R$. As a result the string entropy gives the entropy of the BTZ black hole, $S \sim 2 \pi \sqrt n \sim 2 \pi E_R$ which agrees with eq. (37).
Seen from $r \to \infty$ the string energy is redshifted to
$$E \sim {r_0 \over L^2} E_R \sim M \eqno(44)$$
Since entropy is a number it remains the same and therefore we find that the string tension also gets renormalized to
$$T \sim {GM \over L^2} \eqno(45)$$
We see that the Hagedorn temperature of the string with the above tension gives the Hawking temperature $T_{Hag} \sim (GM)^{1/2}/2 \pi L \sim T_{BTZ}$.

The entropy of the BTZ black hole has been obtained by counting states in the boundary CFT[\CON,\STR]. In that approach, the black hole corresponds to a CFT state with
central charge $c \sim L/G$ and eigenvalue $L_0 \sim  n \sim ML$. The string on the stretched horizon has $\sqrt n \sim L (G/M)^{1/2}$ with
central charge $c \sim O(6)$. It has length $ \sim E_R \ell_s$ and corresponds to a long string which wraps around the boundary $\sim \ell_s/G \sim 1/g^2$
times.

\medskip
\centerline{\bf 6. Conclusions and Discussion}
\medskip

In this paper, we showed that the gravitational entropy of any black object can be understood as the entropy of a highly excited string on the
stretched horizon. For an asymptotic observer at infinity the string has a gravitationally renormalized tension due to the large redshift near the horizon.
The Hawking temperature of the black object is given by the Hagedorn temperature of the string with the renormalized tension.
This result seems to be completely general; it holds for all black objects in any dimension. Therefore it constitutes
a universal explanation of gravitational entropy due to horizons. The (zero or) nonzero entropy of extreme branes and black holes are also obtained as
smooth limits of the nonextreme entropies. As concrete examples, we considered 3 and 5--branes, dyonic strings, M2 branes,
$D=4$ Reissner--Nordstrom black holes and BTZ black holes.
The case of the M--branes are especially
puzzling since the description in terms of strings should not be applicable to them. We stress that, we can only obtain the
the correct scaling of the entropy up to a multiplicative constant in each case. It is hard to imagine how this can be improved without any
supersymmetry or any deeper understanding of why our prescription works.

The complete generality of our results indicate that highly excited strings on stretched horizons describe the fundamental physics on the horizon.
Since the entropy of a string is
proportional to its length, we can view each string bit of length $\ell_s$ as the fundamental degree of freedom on the horizon. Thus, entropy simply counts the
number of string bits on the stretched horizon. The nature of the string is not clear to us. For example, it behaves like a free (or weakly coupled) string
even though naively it seems to be strongly coupled due to the very large number of degrees of freedom. In addition, we could not say much about the central charge
of the string. However, in ref. [\HRS], using the equivalence of power emission and horizon area for black holes, it was shown that one can identify
strings with black holes only if $c=6$. There is every reason to think that this applies to all black holes considered above.
This would mean that the string on the stretched horizon is noncritical.

There seem to be a number unexplained facts about our prescription. First, as mentioned above we use the perturbative string entropy and mass formulas
without any justification. Second, the string describes all the entropy of the black object but only the mass beyond extremality. So the prescription works
only if the extreme mass does not get renormalized even though there is no overall supersymmetry. We do not understand why the two components of mass should
behave so differently.

In our view, one should take our results as ``experimental evidence'' for the prescription used. The prescription seems quite ad hoc but it seems to work
in all cases. If one believes the ``evidence'', then the next step would be to try to understand why this seemingly strange prescription works so well.
The unifying feature of all the horizons discussed above is the fact the the near horizon geometry reduces to Rindler space. Thus it seems that understanding
strings in Rindler space is quite important. The extreme black holes with horizons which are not described by Rindler space are obtained as limits of the
corresponding nonextreme cases.

The universal nature of Bekenstein--Hawking entropy needs a universal microscopic explanation. Strings which live on the stretched horizon with renormalized tension
seem to provide such a universal explanation for gravitational entropy. In fact, to obtain the entropy all one needs to know is the dependence of the Rindler time
on the asymptotic time and mass of the black hole, $\tau_R(t,M)$. This gives the Rindler energy which is the entropy and also the renormalized string tension due to the
redshift. Strings are crucial for microscopically explaining the relation $S=2 \pi E_R$. The energy of the string on the stretched horizon is
$E_{sh} \sim E_R/\ell_s$. On the other
hand the string energy is given by $E \sim S/\ell_s$ which gives $S \sim E_R$.

As far as we know, the only other attempt to give a universal explanation for black hole entropy was made in ref. [\CAR] (See also [\SOL]).
It is interesting that in that approach the crucial ingredient is
the group of diffeomorphisms in the $t-r$ plane for any black hole in any dimension. Our prescription also uses only the properties of the metric in the $t-r$ plane.
It would be interesting to see if these two approaches are related at a deeper level.



\vfill

\refout

\end
\bye